\documentstyle[prl,aps,epsfig,floats,twocolumn]{revtex}
\def\ni{\noindent}

\def\bM{\vec{M}}

\def\bff{\vec{f}}

\def\bdelta{\vec{\delta}}  
\def\bphi{\vec{\phi}}
\def\brho{\vec{\rho}}

\def\hP{{\hat P}}

\def\hsig{{\hat \sigma}}

\def\sig{{\sigma}}

\begin{document}
\draft


\title{Stress transmission and isostatic states of non-rigid particulate systems}
\author{Raphael Blumenfeld}

\address{$^1$ Biological and Soft Systems, Cavendish Laboratory, Madingley Road, Cambridge CB3 0HE, UK} 
\maketitle 
\date{\today} 
\maketitle 

\begin{abstract} 

The isostaticity theory for stress transmission in macroscopic planar particulate assemblies is extended here to non-rigid particles.  It is shown that, provided that the mean coordination number in $d$ dimensions is $d+1$, macroscopic systems can be mapped onto equivalent assemblies of perectly rigid particles that support the same stress field.  The error in the stress field that the compliance introduces for finite systems is shown to decay with size as a power law. This leads to the conclusion that the isostatic state is not limited to infinitely rigid particles both in two and in three dimensions, and paves the way to an application of isostaticity theory to more general systems.
 
\end{abstract}
\narrowtext

Much attention has been given lately to particulate systems both due to their overwhelming technological importance and the
fundamental theoretical challenges that they pose \cite{GranRev}. In particular the micro- and macro-mechanics have focused
research activity following experimental \cite{ForceChainsExp}\cite{PhotoElastic} and numerical \cite{ForceChainsNum}
observations of nonuniform stress fields \cite{Arches}. Specifically, stresses frequently appear to be supported by
arch-like regions, termed force chains, that cannot be straightforwardly described by conventional approaches
\cite{sandpilesmin}.  It has been recognized that to understand this phenomenon it is essential to first understand
transmission of stresses in 'isostatic' systems \cite{Arches}.  Isostatic states are configurations of particles where the
interparticular contact forces are {\it statically determinate}, i.e. from the mechanical equilibrium conditions of balance
of force and torque moments. This means that the interparticular forces can be determined without reference to compliance
and hence to stress-strain relations. Isostatic states are characterized by low mean coordination numbers per particle which
depend on the dimensionality of the system and on the particles roughness. For rough and infinitely rigid particles in
$d$-dimensional systems ($d=2,3$) this number is $z_c = d + 1$, for smooth infinitely rigid particles of arbitrary shape
$z_c = d(d + 1)$ \cite{Bai}\cite{EG}\cite{BEB}, and for smooth infinitely rigid spheres $z_c = 2d$. Isostatic packings of
particles are marginally rigid and such states have been shown to be easy to approach experimentally \cite{BEB}, making them
interesting more than only theoretically. Several empirical \cite{MEC}\cite{TkWi} and statistical \cite{Arches}\cite{EG}
models have been proposed for the macroscopic stress field equations in these systems, suggesting a linear coupling between
the components of the stress tensor. This has been recently established from first principles in the two-dimensional case
for systems of infininitely rigid particles \cite{BaBl}. The new isostaticity theory (IT)  closes the stress field equations
with a constitutive relation between the stress tensor $\hsig$ and a rank-two symmetric fabric tensor $\hP$ which
characterizes the local microstructure:

\begin{equation}
p_{xx}\sig_{yy} + p_{yy}\sig_{xx} - 2p_{xy}\sig_{xy} = 0 \ .
\label{eq:Ai}
\end{equation}
On the scale of a few particles, this equation is a local manifestation of the torque balance condition beyond the global
requirement that $\hsig=\hsig^T$ \cite{BaBl}. 
It then transpired that the coarse-graining of eq. (\ref{eq:Ai}) is not trivial, but this was eventually resolved, making it
applicable for macroscopic systems, albeit with a subtle difference in the interpretation of the constitutive field $p_{ij}$ 
\cite{Bl0405}. This paved the way to several results, most notably it enabled a derivation of the general solution for the
stress field in two-dimensional isostatic granular packings \cite{Bl0408}. The solution turned out to indeed give rise to
force chains and arches. This, not only gave a firm theoretical basis that explains the experimentally observed force
chains, but also provided a way to predict the trajectories of individual force chains. Using these predictions made it
possible to test the theory by direct comparison with experimental measurements.

However, much controversy surrounds the validity of the new theory. In particular, because it has been developed for 
infinitely rigid particles there remained questions concerning its validity to general particulate systems, whose rigidity
is unavoidably finite. The clarification of this point is a crucial first step towards bridging between IT and elasticity
theory. A detailed examination of this issue, both in two and in three dimensions, is the aim of this paper.  

\begin{figure} 
\centerline{\psfig{file=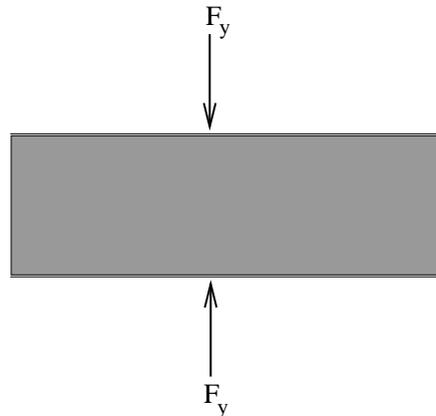,height=5.5cm}} 
\caption{ 
The loading on a packing of grains by a force $F_y$ that is distributed evenly on the shown surfaces} 
\end{figure} 

\bigskip

\ni{\bf Two dimensions}: 

Consider a polydisperse planar packing of $N$ particles of arbitrary shape and typical area $a$ 
\cite{polydisperse}, confined to within a square container of dimensions $L^2 \sim aN $. All the particles are presumed to be made of the same material whose elastic properties are known \cite{DiffMat}. The packing is loaded by an infintesimally small external compressive force $F_y$ on two opposite boundaries, as shown in figure 1. The load compresses the particles against one another slightly and the contacts between neighbouring particles consist of short lines. The line contacts, rather than point contacts, between particles constitute the main difference between packings of compliant and infinitely rigid particles. The criterion for 'smallness' of $F_y$ is that the contact lines are smaller that the linear size of the corresponding particles \cite{smallness}. For a system of infinitely rigid particles to be statically determinate in two dimensions the mean coordination number per particle must be $z_c = 3$ up to a boundary-to-bulk correction term. We wish to determine whether stress fields that develop in assemblies of compliant particles that satisfy this condition are also governed by the equations of IT.

An ideal resolution of the issue would be to establish whether {\it all} the interparticular forces can be determined, at
least in principle, from balance conditions alone. If this is possible then the system is statically determinate and
isostaticity theory must apply. The main difference between the geometries of infinitely rigid and compliant systems is 
that while in the former the interparticular forces act at a point of contact, in the latter they are continuously
distributed along contact lines.
Let us examine the contact between two touching particles, $g$ and $g'$. The particles press on one another
with a force that is distributed along the contact line with density $\phi(x)$, where $x$ is a length parameter that varies
from $0$ to $l$ along the line (see figure 2). Due to the arbitrary shapes of particles this force density need not be
uniform. Both a force and a torque moment are transmitted through the contact and these are given by

\begin{equation}
\bff^{gg'} = \int_0^l \bphi(x) dx
\label{eq:Aii}
\end{equation}
and 
\begin{equation}
\bM^{gg'} = \int_0^l \bphi(x) \times \brho(x) dx \ .
\label{eq:Aiii}
\end{equation}
Here $\brho(x)$ is the position vector from the centroid of the particle (defined as the mean position vector of the contact points of the particle $g$, see figure 2). 

\begin{figure} 
\centerline{\psfig{file=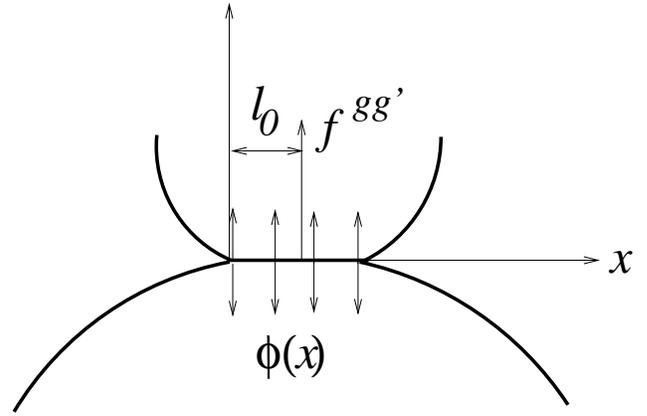,height=5.5cm}} 
\caption{ 
The distribution of forces, $\bphi(x)$ along the contact line between grains $g$ and $g'$. The contact line is parameterised by $0<x<l$ from left to right. The mean force is $\bff^{gg'}$ located at a distance $x=l_0$, found from the first and second moments of the force density} 
\end{figure} 

So, do we need to determine the entire force distributions along the contact lines? Considering relations (\ref{eq:Aii}) and
(\ref{eq:Aiii}), the answer is encouragingly no.  The torque moment can be represented by a single force of magnitude
$\bff^{gg'}$ acting at a point $x=l_0$ that lies between $x=0$ and $x=l$ (it is straightforward to see that $l_0$ cannot be
outside this section), and whose location is determined by the relation

\begin{equation}
\bff^{gg'} \times \brho(l_0) = \bM^{gg'} \ .
\label{eq:Aiv}
\end{equation} 
Thus, it seems that, at least in principle, we can reduce the problem to find the discrete forces $\bff^{gg'}$. In
mechanical equilibrium these interparticular forces balance out \cite{NoBodyForces}

\begin{equation}
\sum_{g'} \bff^{gg'} = 0 \ .
\label{eq:Av}
\end{equation}
The stress field can be defined in terms of the force moments around the particles 

\begin{equation}
S^g_{ij} = \sum_{g'} f^{gg'}_i \rho^{gg'}_j(l_0) \ .
\label{eq:Avi}
\end{equation}
The torque balance condition for every particle amounts to the requirement that $S^g_{ij}=S^g_{ji}$. With this definition
the stress within a given region inside the material is the area average of the force moments over the particles within the
region.

Expressions (\ref{eq:Aii}), (\ref{eq:Aiii}) and (\ref{eq:Avi}) suggest that the stress field is determined only by the
forces $\bff^{gg'}$, rather than by the entire distributions of the contact forces. It follows that if we knew the locations of the points $l_0^{gg'}$ where the equivalent interparticular forces act then we could map the system of compliant particles onto an equivalent one of infinitely rigid and infinitely rough particles that contact at these points, as illustrated in figure 3. The equivalent system would have the same mean coordination number and it would transmit the same interparticular forces. Therefore, it would also have the same macroscopic stress field. Since the stress field in the equivalent packing is governed by IT then this would lead to the important conclusion that the original packing of compliant particles is also isostatic and is indeed described
by IT.

\begin{figure} 
\centerline{\psfig{file=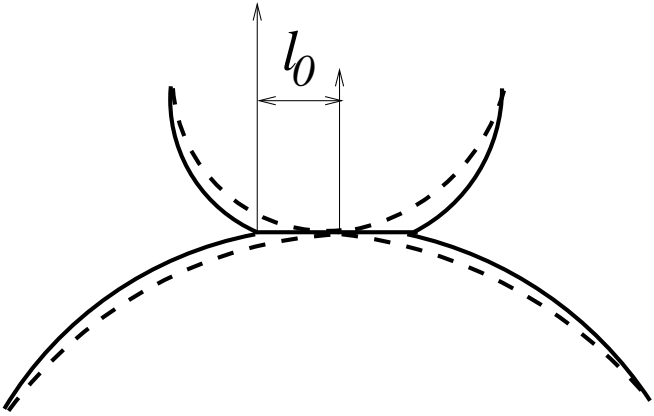,height=5.5cm}} 
\caption{ 
The equivalent system of ideally rigid particles (dashed lines) touch at the point $x=l_0$ along the contact line of the original compliant particles (full lines)} 
\end{figure} 

However, there still remains the issue of the adaptation of the formalism to the structure of compliant systems. In
particular, recall that IT relies on the identification of the geometric tensor whose components $p_{ij}$ depend directly
on the positions of the contact points. Thus, the question that we are faced with is whether it is possible to identify the
points along the contact lines, $l_0^{gg'}$.
Here we appear to have a problem. To determine the locations of these points requires using relations (\ref{eq:Aiii}) and
(\ref{eq:Aiv}), which in turn require full knowledge of $\bphi(x)$ at the contact between every two particles. But this is
tantamount to a solution of the interparticular forces in the first place. Does this mean that we cannot find the equivalent
rigid packing? Have we reached a dead end?

Not necessarily. The conundrum can be resolved as follows. Let us introduce a judiciously chosen approximate equivalent 
system for which we can use IT to solve for the stress field. The idea is to show that the difference between the 
approximate and the true fields diminishes as $N^{-\alpha}$ ($\alpha > 0$) when the system size increases and therefore that
the approximate solution converges to the true solution for macroscopic systems. The equivalent packing is generated by
choosing the forces $\bff^{gg'}$ to act at the centres of the contact lines. This requires only knowledge of the structure,
not the force distributions. Let us construct the geometric tensor $\hP$ for the equivalent system, using the definition in
\cite{BaBl} and coarse-grain it using the procedure in \cite{Bl0405}. Together with the boundary data, we can now determine
the stress field using the solution of reference \cite{Bl0408}.

The deviation of this solution from the 'true' stress field arises from the error in the position vectors that point from
the centroids of the grains to the location of $\brho^{gg'}(l_0)$. Defining the error vectors as $\bdelta^{gg'} =
\brho^{gg'}(l/2) - \brho^{gg'}(l_0)$, the error in the stress field around particle $g$ is given by

\begin{equation}
\delta \sig^g_{ij} = \frac{1}{a^g} \sum_{g'} \delta^{gg'}_i f^{gg'}_j \ ,
\label{eq:Avii}
\end{equation}
where $a^g$ is the area associated with particle $g$. 

Now, a continuous stress field representation is only useful on scales that contain a good number of particles $M$ but where
$M \ll N$. For macroscopic description, the system is regarded as a collection of {\it continuous} such units. The error
made in the stress within such a unit region of area $A^{\Gamma} = \sum_g a^g \sim aM$ is

\begin{eqnarray}
\delta\sig_{ij}^\Gamma\, & = & \delta\left[ \frac{1}{A^\Gamma} \sum_g a^g \sig_{ij}^g \right]  \nonumber \\
& = & \frac{1}{A^\Gamma} \sum_g a^g \delta\sig_{ij}^g = \frac{1}{A^\Gamma} \sum_{g,g'\in\Gamma}\delta_i^{gg'}f_j^{gg'} \ .
\label{eq:Aviii}
\end{eqnarray}
The error in the stress field is linear in the $\delta^{gg'}$ and its magnitude depends directly on the correlations between these quantities;

\begin{equation}
\left( \delta\sig_{ij}^\Gamma\right)^2 = \frac{1}{\left(A^\Gamma\right)^2}\sum_{g,g', g'',g'''\in\Gamma} 
\left( \delta_i^{gg'} \delta_i^{g''g'''} \right) f_j^{gg'} f_j^{g''g'''} \ ,
\label{eq:Aviiia}
\end{equation}
Let us now assume that in isotropic packings the error vectors $\delta^{gg'}$ are random and uncorrelated. This, of course,
may not be the case since correlations may arise from the history of the dynamics that gave rise to the structure, as well
as from inhomogeneities in material properties and granular characteristics that lead to nonlinear contact lines. However,
in the absence of evidence to the contrary it is plausible that under small loading these effects are negligible. Then 
the sum in (\ref{eq:Aviii}) can be regarded as a two-dimensional Markovian random walk and it increases as $O(M)$, up to
logarithimic terms. But the area also increases as $O(M)$ and hence the entire expression decreases as $1/M$. We therefore
arrive at the conclusion that

\begin{equation}
\delta\hsig^\Gamma \sim \frac{1}{\sqrt{M}} \ .
\label{eq:Aix}  
\end{equation}
This result encouragingly support the idea that the approximate and the true stress fields converge in the macroscopic limit. But we are not finished yet. Having partitioned the system into $N/M$ basic units of $M$ grains, we now face the acid test of the analysis. We need to determine the size of the discrepancy between the boundary data and the corresponding data derived from the approximate field. Using the same rationale, it is assumed that in isotropic systems the errors in the stresses (\ref{eq:Aix}) in different basic units are independent. It follows that the error in the stress field at the boundary (which is normalised by the total area) is of order $O(\sqrt{M/N}) \ll 1$. For macroscopic systems this error is indeed negligibly small. We have therefore reached the desired result; in macroscopic packings of compliant grains the approximate and the true stress fields are the same and can be obtained by solving for the isostatic stress in the equivalent infinitely rigid packing.

To make the analysis even more quantitative, let the particles' Young modulus be $E$ and let us assume that their local radius of curvature is typically $R = \alpha\sqrt{a}$. In isotropic systems it is expected that the value of the parameter $\alpha$ would be distributed over particles around $1/\sqrt{\pi}$. For monodisperse circular particles the distribution of $\alpha$ is almost a $\delta$-function around this value. A sensible choice of $M$ is such that there are many particles in a unit region on the boundary that are pressed by the boundary loading. With this choice the fluctuations of the force on the boundary particles can be disregarded and the mean force per particle in the $y$-direction is $F_y/\sqrt{N}$. Two particles in contact exerting a normal force $f_n$ on one another deform slightly and according to Hertz theory the line contact between them is

\begin{equation}
w = 4\sqrt{\frac{f_n R_{e}}{\pi E'}} \ .
\label{eq:Ax} 
\end{equation}
long. In this expression $R^e$ is an effective radius, $1/R^e = 1/R^g + 1/R^{g'}\approx 2\sqrt{\pi/a}$, $E' = 2(1-\nu^2)/E$
and $\nu$ is Poisson's ratio. Substituting for the compressive force $f_n$ gives that the width of a typical line contact is

\begin{equation}
w \approx \left(\frac{4}{\pi}\right)^{3/4}\sqrt{\frac{F_y}{E'}} \left(\frac{a}{N}{1/2}\right)^{1/4} \ .
\label{eq:Axi}
\end{equation}
The error in the distance between the middle of the line and the true point $l_0$ is at most $w/2$. Thus, using relations (\ref{eq:Avii}) and (\ref{eq:Axi}), and taking into consideration that there are on average three contacts per particle, the typical error made in the computation of the stress around any one particle is bounded by

\begin{equation}
| \delta \sig^g_{ij} | < \frac{3wf_n}{2a} \approx \frac{3\sqrt{2}}{\pi^{3/4}}\sqrt{\frac{F_y^3}{E'}}\left(aN\right)^{-3/4} \ .
\label{eq:Axii}
\end{equation}
This calculation gives the precise power $\alpha=3/4$ with which the error between the stress fields decays with the number of particles $N$ and completes the proof.

Recalling that theories for ideal rigid particles predict stresses that propagate nonuniformly along arches \cite{Arches} \cite{MEC} \cite{Bl0408}, this explains why such force chains are also observed in packings of compliant particles \cite{ForceChainsExp}-\cite{ForceChainsNum}. Moreover, since the trajectories of force chains can be predicted in ideal packings \cite{Bl0408} then the above suggests that these predictions can be extended to systems of compliant particles. It would be interesting to compare these predictions with the actual trajectories observed in realistic systems, such as those of reference \cite{ForceChainsExp}.

\bigskip

{\bf Three dimensions}:

\ni The discussion of the three-dimensional systems follows the same rationale. The systems considered here consist of compliant and non-slipping particles, slightly compressed under a low external load. The mean
coordination number is four per particle and the contacts make small two-dimensional surfaces. We assume, for simplicity, that the particles have homogeneous elastic properties, in which case the contact surfaces are planar. It is straightforward to lift this assumption and extend the results to systems of particles with nonuniform properties. The difference between such a system and one of infinitely rigid particles is that the interparticular forces are distributed across the contact surfaces.
The contact surface between particles $g$ and $g'$ can be described by by a position vector $\brho^{gg'}(x,y)$, where $x$ and $y$ parameterise the surface of the contact. Terming the interparticular force density
$\bphi^{gg'}(x,y)$, the mean force between the two particles is

\begin{equation}
\bff^{gg'} = \int_s \bphi^{gg'}(x,y) dxdy \ ,
\label{eq:Bi} 
\end{equation}
where $s$ stands for an area integration across the contact surface. The mean torque moment on the surface is

\begin{equation}
\bM^{gg'} = \int_s \bphi^{gg'}(x,y) \brho^{gg'}(x,y) dxdy \ .
\label{eq:Bii} 
\end{equation}
From eqs. (\ref{eq:Bi}) and (\ref{eq:Bii}) we now extract a position vector $\brho^{gg'}_0$ by using 

\begin{equation}
\bff^{gg'}\times\brho^{gg'}_0 = \bM^{gg'}\ .
\label{eq:Biiii}
\end{equation}
It is straightforward to verify that if the contact surface is planar then $\brho^{gg'}_0$ corresponds to a point on the surface
$\brho^{gg'}_0\equiv\brho^{gg'}(x_0,y_0)$. Non-planar surfaces, which may result from non-uniform elastic properties in the particles, do not
pose a limitation on this analysis as long as the deviation from the plane is smaller than the size of either of the particles in contact. 
We can now define the equivalent ideal system by postulating that its infinitely rigid  particles make contacts at the points $\brho^{gg'}_0$.

This done, we face the same conundrum as in the two-dimensional case; the determination of the contact points of the equivalent system poses the same level of difficulty as the original determination of the contact force distributions. For the resolution of this problem we follow the same logic as before. We construct an approximate system, for which isostaticity theory can be applied to determine the stress field, and then we show that the approximate field converges to the true field as the size of the system increases. 

The contact points of the approximate system of infinitely rigid particles are postulated to be the centroid of the contact surfaces of the true system of compliant particles,

\begin{equation}
\brho^{gg'}_{approx} = \frac{\int_s \brho^{gg'}(x,y) dxdy}{\int_s dxdy}
\label{eq:Biv}
\end{equation}
These points are well defined from the geometry. The stress field in the approximate ideal packing of rigid particles is 

\begin{equation}
\sig^g_{ij} = \frac{1}{V^g}\sum_{g'} f^{gg'}_i \left(\rho^{gg'}_{approx}\right)_j \ ,
\label{eq:Bv}
\end{equation}
where $V^g$ is the volume associated with particle $g$. There are several ways to define the volume $V^g$ such that $V^{sys}=\sum_g V^g$, but the precise definition is not essential for the present discussion.  The error in the stress around particle $g$ originates from the deviations of the true positions of the effective forces from the approximate positions $\delta\brho^{gg'}=\brho^{gg'}_{approx}-\brho^{gg'}_0$, 

\begin{equation}
\delta\sig^g_{ij} = \frac{1}{V^g}\sum_{g'} f^{gg'}_i \delta\rho^{gg'}_j \ .
\label{eq:Bvi}
\end{equation}
Consider a region $\Gamma$ containing $M\ll N$ particles. The volume of the region is of order $V^\Gamma = \sum_g V^g \approx VM$, where $V$ is the typical particle volume. The error in the stress over this region is 

\begin{eqnarray}
\delta\sig^\Gamma_{ij}\, &= \delta\left[\frac{1}{V^\Gamma}\sum_{g} V^{g} \sig^{g}_{ij} \right] \nonumber \\
&= \frac{1}{V^\Gamma}\sum_{g,g'\in\Gamma}\delta\rho^{gg'}_i f^{gg'}_j \ .
\label{eq:Bvii}
\end{eqnarray}

Following a similar analysis as in two dimensions we end up with a three-dimensional Markovian random walk of the vectors $\delta\brho^{gg'}$. Assuming now that in isotropic systems these are uncorrelated (at least above some lengthscale), we conclude that $\left(\delta\sig^\Gamma_{ij}\right)^2$ increases as $O(M^{1/2})$ and therefore that the error decreases as $O(M^{-1/2})$. By considering many such regions we are then led to the conclusion that the error at the boundary between the true and approximate stress fields decreases again as $O(\sqrt{M/N})\sim O(L^{-3/2}) \ll 1$.

We can make the analysis again more quantitative by taking into consideration the elasticity of the particles, their typical size and assuming Hertzian interaction. Following the line of reasoning as that leading to eq. (\ref{eq:Axii}), we find that according to Hertz theory the diameter of the area of contact between two grains is 

\begin{equation}
w = 2\left(\frac{R^ef_n}{K}\right)^{1/3} \ ,
\label{eq:Bvii} 
\end{equation}
where $f_n$ is the force pressing them together, $1/K\equiv (3/4)[(1-\mu^2)/E + (1-\mu'^2)/E']$ is an effective elastic constant, and $1/R^e=1/R^g + 1/R^{g'}$ is an effective curvature. Taking a cube of $N$ particles and pressing on one of its surfaces by a force $F_z$ gives that on average per grain there is a normal force of order $F_z/N^{2/3}$. Recalling that there are on average four contacts per particles, the error in the stress around a particle can be bounded by 

\begin{equation}
| \delta \sig^g_{ij} | < \frac{2wf_n}{V} \approx \frac{3}{\pi \left(R^e\right)^{8/3}K^{1/3}} F_z^{4/3}N^{-8/9} \ .
\label{eq:Bviii}
\end{equation}

Thus we have demonstrated that, just as in two dimensions, the isostatic solution for the equivalent system of ideally rigid particles converges to the true stress field of the packing of compliant particles as the size of the system increases. 

To conclude, it has been shown in this note that isostaticity theory is not limited to packings of infinitely particles. Rather, this theory acn be used to describe stress fields in macroscopic systems of compliant particles. The only condition that such systems must satsify is the same as the one for rigid-particle isostatic systems; that the number of contacts per particle is $z_c = d+1$ in $d=2, 3$ dimensions. 

The next step towards bridging between isostaticity and elasticity theories involves considering packings where the mean coordination number is larger than $z_c$. A suggestion in this direction has been made by this author \cite{Bl0408} and a detailed formulation of an isoelasticity theory will be reported elsewhere.

\vspace{0.5in}

\ni {\bf Acknowledgements} 
 
\ni I am grateful to Prof. Robin Ball for critical comments.


\begin{thebibliography}{99}
\bibitem{polydisperse} A typical grain size is defined here for convenience only; there is no loss of generality and the discussion is equally valid for packings of particles with any distribution of grain sizes.
\bibitem{DiffMat} The extension of the analysis presented here to packings of particles of different materials is
straightforward
\bibitem{smallness} It should be noted that for finite systems it is always possible to find a finite small force $F_y$ that satisfies this criterion. However, the magnitude of this force decreases with system size
\bibitem{NoBodyForces} Without loss of generality, the packing is presumed to experience no body forces
\end{thebibliography}
\end{document}